\documentclass{ws-mpla}

\begin{document}

\markboth{W.S. Hou} {
Highlights from Five Years at B Factories}

%
\catchline{}{}{}{}{}
%

\title{Highlights from Five Years at the B Factories
}

\author{\footnotesize Wei-Shu Hou
}

\address{Department of Physics, National Taiwan University\\
Taipei, Taiwan 106, R.O.C.
\\
wshou@phys.ntu.edu.tw
}



\maketitle

\pub{Received (Day Month Year)}{}

\begin{abstract}
The highlights and conflicts at the B Factories are briefly
reviewed. $CP$ violation was established in 2001 in $B^0 \to
J/\psi K_S$ and related modes, which has now become a precision
measurement of $CP$ violation in $B^0$-$\overline B^0$ mixing.
However, the situation for the $B^0 \to \pi^+\pi^-$ and charmless
$b\to s$ modes, which probe also CP violation in the {\it decay}
amplitude, are not quite settled yet. They could be hinting at
presence of both strong ($CP$ conserving) and new physics ($CP$
violating) phases. We critically assess the developments and
discuss some related discrepancies and highlights, such as
observation of direct $CP$ violation, and make a projection
towards the next few years.

\end{abstract}

\ccode{PACS Nos.: 13.25Hw, 11.30Er, 12.15Hh, 14.40Nd}

\section{Introduction}

$CP$ violation (CPV) was discovered\cite{eK} in $K^0$-$\overline
K^0$ mixing in 1964. With it we came to realize, in terms of the
Sakharov conditions,\cite{Sakharov} that CPV is a prerequisite for
understanding the baryon asymmetry of our Universe. However, it
took another 35 years, until the measurement\cite{epK} of
$\varepsilon^\prime_K$, or ``direct" CPV (DCPV), for us to put the
superweak\cite{SW} (CPV in $K^0$ mixing only, but not in $K$
decay) model to rest. Thus was the paucity of CPV in the last
century.

The 21st century began with a roaring start in observations of CPV
in the $B$ meson system. Indirect CPV, or CPV in $B^0$-$\overline
B^0$ mixing, was firmly established\cite{beta01,phi101} in 2001.
By 2004, DCPV was established\cite{AKpi1,AKpi2} in $B^0\to
K^+\pi^-$ decay. The history of the kaon system was repeated, in
not quite the same way, in just 3 years. The $B$ system also opens
up a host of CPV and other observables.

The precursor to the modern view of CPV came a year before 1964,
with the Cabibbo angle ($\sin\theta_C$) proposal\cite{Cabibbo}
that unified strange and nonstrange weak decays. By 1970, the GIM
mechanism called for two generations of quarks and leptons, making
$\sin\theta_C$ a genuine rotation angle of a $2\times 2$ matrix.
The two generation picture was completed with the $J/\psi$
discovery of November 1974. But in 1973, Kobayashi and Maskawa
(KM) realized\cite{KM} that, upon generalizing quark mixing to 3
generations, i.e. from $2\times 2$ to $3\times 3$ matrix, one has
a {\it unique} CPV phase. Together with the establishment of the
gauge theory of strong and electorweak interactions, by the
mid-1970's the CKM quark mixing picture became an integral part of
the Standard Model (SM), which has withstood test upon test for
the past 30 years.

The 
$\varepsilon_K^\prime$ parameter suffers from hadronic
uncertainties that make the extraction of fundamental parameters
difficult. In 1979 it was realized\cite{CSBS} that the $B$ system
offers much better prospects. In the so-called mixing-decay CPV
mechanism in $B^0 \to J/\psi K_S$, not only one expects the effect
to be large, but because the decay amplitude is free from CPV
phases, one can make a clean measure of the CPV phase
$\sin2\phi_1$ (also called $\sin2\beta$) in the CKM ``unitarity
triangle".

Several developments were pivotal to the realization of such
measurements.
In 1983, the $B$ hadron lifetime was found to be much
prolonged,\cite{tauB} and $b\to c$ transitions dominated over
$b\to u$ transitions. This stimulated the application of silicon
based vertex detectors, while we now know that, taking $V_{us}
\equiv \lambda \simeq 0.22$ as real, we have
\begin{equation}
V_{cb} = A\lambda^2, \ \ \ V_{ub} = A\lambda^3(\rho-i\eta),
\label{eqn:vcbvub}
\end{equation}
with $A\simeq 0.8$ and $\sqrt{\rho^2+\eta^2}\sim 0.3$--0.4. It is
remarkable that the progressive smallness of off-diagonal CKM
matrix elements explains why CPV effect is so small in SM, as one
needs the participation of all 3 generations.
Second, in 1987 the ARGUS experiment discovered\cite{ARGUS} large
$B^0$-$\overline B^0$ oscillations, i.e. $\Delta m_{B_d} \simeq
0.5 \Gamma_B$. This was not only the harbinger for the heaviness
of the top quark, it provided an almost ideal setting for the
mixing-decay CPV mechanism to be realized.
Finally, as the CLEO experiment was making upgrades, and when
discussions were ongoing at PSI for a new $B$ facility, Oddone
suggested in 1988 to make the $e^-$ and $e^+$ beams {\it
asymmetric} in energy. The boosted $B$ mesons made time-dependent
measurements possible. Serious studies soon followed at KEK and
SLAC, and by 1994 both places embarked on the construction of
asymmetric $e^+e^-$ ``B factories", the very successful KEKB and
PEP-II colliders, as well as the Belle and BaBar detectors. After
commissioning in 1999, by 2004 each experiment had accumulated
more than 200M $B\overline B$ events, more than a factor of 20
over what CLEO collected throughout the 1990's.

The aim of this brief review is to give an account of the
competitive history, the major milestones, as well as the ongoing
debates if not controversies.

\section{Raison D'Etre:  $B^0 \to J/\psi K_S$ and $\sin2\phi_1/\sin2\beta$ Measurement}

The physics of CPV in mixing-decay interference is rather close to
the classic double slit experiment. Consider a $CP$ eigenstate $f$
that both $\overline B^0$ and $B^0$ can decay to. Besides the
$\overline B^0\to f$ decay amplitude, an initial $\overline B^0$
meson can oscillate into a $B^0$ meson and then decay to $f$. The
interference pattern is measured to determine CPV in both the
mixing and decay amplitudes. We note that CPV is measurable only
when the two interfering amplitudes, $A_1$ and $A_2$, have both
$CP$ violating as well as $CP$ conserving relative phase
differences. That is, 
\begin{equation}
A_{CP} =
\frac{2|A_1||A_2|\,\sin(\theta_1-\theta_2)\sin(\delta_1-\delta_2)}
     {|A_1|^2+|A_2|^2+2|A_1||A_2|\,\cos(\theta_1-\theta_2)\cos(\delta_1-\delta_2)},
 \label{eqn:ACP}
\end{equation}
which vanishes if either the $CP$ violating or $CP$ conserving
phase differences $\theta_1-\theta_2$, $\delta_1-\delta_2$ vanish.
Here, the oscillation phase $e^{-i\Delta m\,t}$ provides the
latter.

For the ``golden mode" of $f= J/\psi K_S$, the $\overline B^0 \to
J/\psi K_S$ decay is dominated by the tree level $b\to c\bar c s$
hence $\propto V_{cb}V_{cs}^*$ in amplitude. Thus, to a very good
approximation, the decay amplitude is real, and the mixing-decay
mechanism measures the CPV phase in the $\overline B^0 \to B^0$
mixing amplitude, which is $\propto V_{td}^{*2}$ in SM.

The CKM quark mixing matrix $V$ governs the strength of $d_j \to
u_i$ weak transitions. With Eq.~(\ref{eqn:vcbvub}), it can be put
in the form
\begin{equation}
 V =
    \left( \begin{array}{ccc}
    V_{ud}  &   V_{us}  &  V_{ub}  \\
    V_{cd}  &   V_{cs}  &  V_{cb}  \\
    V_{td}  &   V_{ts}  &  V_{tb}  \end{array} \right)
 \simeq
    \left(  \begin{array}{ccc}
    1 - \lambda^2/2               &     \lambda     &  A\lambda^3(\rho - i\eta) \\
      - \lambda                   & 1 - \lambda^2/2 &  A\lambda^2 \\
    A\lambda^3(1 - \rho - i \eta) &   -A\lambda^2   &  1
   \end{array} \right),
\label{eqn:CKM}
\end{equation}
to order ${\cal O}(\lambda^4)$. The matrix $V$ is unitary, and the
relation we probe is
\begin{equation}
 V_{ud}V_{ub}^* + V_{cd}V_{cb}^* + V_{td}V_{tb}^*  = 0,
\label{eqn:UT}
\end{equation}
which is visualized as the unitarity triangle (UT)  shown in
Fig.~1. It is remarkable that the fundamental phenomena of CPV can
have such simple geometric representation. The CPV phase of
$V_{td}^{*2}$ probed by $\overline B^0/B^0 \to J/\psi K_S$ is
$\sin2\phi_1$ (or $\sin2\beta$).

\begin{figure}[t!]
\begin{center}
{\psfig{file=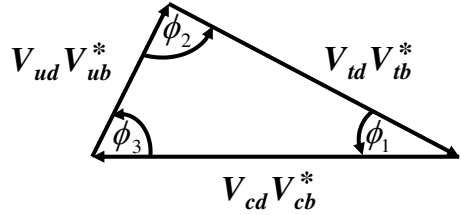,width=4.8in}} \vspace{-6.3cm} 
\end{center} \caption{The unitarity triangle, Eq.~(\ref{eqn:UT}),
probed by the B factories.}
\end{figure}

\subsection{Master Formula and Template for TCPV Measurement}

At B factories, the time-dependent CPV (TCPV) asymmetry for $B \to
f$ decay is
\begin{eqnarray}
 A(\Delta t) & \equiv & \frac{
      \Gamma(\overline{B}^0(\Delta t) \to f) -
      \Gamma(B^0(\Delta t) \to f)}
     {\Gamma(\overline{B}^0(\Delta t) \to f) +
      \Gamma(B^0(\Delta t) \to f)}
\nonumber \\
             & = &
     -\xi_f({\cal S}_f\sin\Delta m\Delta t +
            {\cal A}_f\cos\Delta m\Delta t),
\label{eqn:ACPt}
\end{eqnarray}
where $\Delta m \equiv \Delta m_{B_d}$, $\xi_f$ is the $CP$
eigenvalue of $f$, $B^0(\Delta t)$ denotes the state at time
$\Delta t$ starting from $B^0$ at $\Delta t = 0$, and (BaBar uses
${\cal C}_f \equiv -{\cal A}_f$)
\begin{equation}
 {\cal S}_f = \frac{2\,{\rm Im}\, \lambda_f}{|\lambda_f|^2+1},
  \ \ \
 {\cal A}_f = \frac{|\lambda_f|^2-1}{|\lambda_f|^2+1},
\label{eqn:SfAf}
\end{equation}
where ${\cal A}_f$ measures DCPV, and $\lambda_f$ is defined as
\begin{equation}
 \lambda_f = \frac{q}{p} \frac{\langle f|S|\overline B^0\rangle}
                                {\langle f|S|B^0\rangle},
\label{eqn:lambdaf}
\end{equation}
which depends on both $B^0$ mixing, i.e. $B_{H,L} = p\, B^0 \mp
q\,\overline B^0$, and decay to state $f$. The lifetime difference
between the two neutral $B$ mesons have been ignored (a very good
approximation for $B_d$), so $q/p \cong e^{-2i\phi_1}$ (so
$|q/p|\cong 1$). For the golden $J/\psi K_S$ mode, the decay
amplitude is real in the standard phase convention, hence
\begin{equation}
 {\cal S}_{J/\psi K_S} \cong \sin2\phi_1,
  \ \ \
 {\cal A}_{J/\psi K_S} \cong 0,
\label{eqn:JpsiKs}
\end{equation}
to very good accuracy.
Many other $b\to s(c\bar c)_{\rm charmonium}$ modes are also
collected and, correcting for $\xi_f$, adds to the statistics.

The $J/\psi K_S$ events are collected by detecting $J/\psi \to
\ell^+\ell^-$ ($\ell = e$, $\mu$) and $K_S \to \pi^+\pi^-$ (and
$\pi^0\pi^0$). Two CMS variables that utilize the special
kinematics of $\Upsilon(4S) \to B\overline B$ decay greatly
enhances signal over background events. One is the
beam-constrained mass $M_{bc} = \sqrt{(E_{\rm CM}/2) - \vec p_{\rm
meas}^2}$, and the other is the energy difference $\Delta E =
E_{\rm meas} - E_{\rm CM}/2$, where $\vec p_{\rm meas}$ and
$E_{\rm meas}$ are the measured momentum and energy in CMS.
Knowing that $\Upsilon(4S) \to B\overline B$ only, and
substituting the much better known CMS beam energy $E_{\rm CM}/2$
for $E_B$ greatly improves resolution.

Two special requirements in Eq.~(\ref{eqn:ACPt}) make construction
of the B factories necessary. Since the $J/\psi K_S$ final state
cannot tell between $B^0$ or $\overline B^0$ decay, we need to
``tag" its flavor. At B factories one utilizes the quantum
phenomenon that, after $\Upsilon(4S)$ decay, the $B\overline B$
system remains coherent until one of the $B$ mesons decays.
Assuming that this ``tagging" side is a $B$ decay at $t_{\rm
tag}$, then the other side evolves as a $\overline B^0$ meson
until it decays to the $CP$ eigenmode $J/\psi K_S$ at time
$t_{CP}$. Thus, $\Delta t \equiv t_{CP} - t_{\rm tag}$ in
Eq.~(\ref{eqn:ACPt}). Note that $\Delta t$ can be of either sign.

Since $B$ momentum in $\Upsilon(4S)$ frame is very small, to
observe the decay points of the two $B$ mesons, one boosts the
$\Upsilon(4S)$ frame at an asymmetric B factory. To a very good
approximation $\Delta t \simeq \Delta z/\beta\gamma c$, where
$\beta$ is the boost of $\Upsilon(4S)$ in lab frame, and
$\beta\gamma = 0.56$ and 0.425, respectively, for PEP-II and KEKB.
One also needs a vertex detector (SVT at BaBar and SVD at Belle)
of sufficient accuracy.

For flavor tagging, one utilizes primary $b\to \ell^-X$ and
secondary $b\to c\to \ell^+$ leptons, secondary $K^\pm$ and
$\Lambda$'s from $b\to c\to s$ sequence, low energy $\pi^\pm$ from
$D^{*\pm}$, and high energy tracks such as $\pi$'s from $B\to
\overline D\pi$. The results are combined into a multidimensional
likelihood function to determine a tag-side charge $q=\pm$.
Reconstructed self-tagged modes (such as $D^{*-}\ell^+\nu$,
$D^{(*)-}\pi^+$, $D^{*-}\rho^+$ etc.) from actual data are used to
measure the wrong-tag fraction $w_i$ (leading to a dilution factor
of $1-2w_i$ in Eq.~(\ref{eqn:ACPt})) for each tagging or purity
category $i$. The total effective tagging efficiency is around
0.27--0.28.

One final technicality is ``blind" analysis. Since the $CP$
asymmetry $A(t)$ can be of either sign, the analysis is performed
``blind" to avoid bias. That is, the value of $\sin2\phi_1$ (and
the $CP$ asymmetry in the $\Delta t$ distribution) from the fit
remains hidden until the analysis is completed. The statistical
error is largely unaffected, but all systematic uncertainties can
be studied without knowing the value of $\sin2\phi_1$.

\subsection{Measurements}

After approval in 1993 and 1994, respectively, both the SLAC and
KEK B factories were commissioned successfully and reported
engineering results in 1999. Based on 9 and 6.2 fb$^{-1}$ of data,
respectively, the first measurements for $\sin2\beta$ and
$\sin2\phi_1$ were reported at the ICHEP 2000 meeting in
Osaka,\cite{Osaka}
\begin{eqnarray}
 \sin2\beta &=& 0.12\pm0.37\pm0.09, \ \ \ \ ({\rm BaBar,\ 9\ fb}^{-1})
  \nonumber \\
 \sin2\phi_1 &=& 0.45^{+0.43+0.07}_{-0.44-0.09}, \ \ \ \ \ \ \ \ \ ({\rm Belle,\ 6.2\ fb}^{-1})
\label{eqn:phi1Osaka}
\end{eqnarray}
which, being consistent with zero, gave physicists the impression
that $\sin2\beta$ might deviate\cite{KagNeu} from SM expectations!
This continued to be the case, especially for BaBar, with the
first published results,\cite{phi1beta01}
\begin{eqnarray}
 \sin2\phi_1 &=& 0.58^{+0.32+0.09}_{-0.34-0.10}.
 \,\ \ \  \ \ \ \ \ \ ({\rm Belle,\ 10.5\ fb}^{-1})
 \label{eqn:phi10301} \\
 \sin2\beta &=& 0.34\pm0.20\pm0.05, \ \ \ \
({\rm BaBar,\ 20.7\ fb}^{-1})
 \label{eqn:beta0301}
\end{eqnarray}

Compared to the summer 2004 average of\cite{HFAG}
\begin{equation}
 \sin2\beta/\sin2\phi_1 = 0.726 \pm 0.037, \ \ \ \ ({\rm HFAG\ Summer\ 2004})
\label{eqn:phi104}
\end{equation}
(B factory average is 0.725) BaBar's result is low by $2\sigma$.
Just before Lepton-Photon 2001, however, BaBar
reported\cite{beta01}
\begin{equation}
 \sin2\beta = 0.59 \pm 0.14 \pm 0.05. \ \ \ \ ({\rm BaBar,\ 32M}\ B\overline B)
\label{eqn:beta01}
\end{equation}
Only 9M $B\overline B$ events were added, but with significant
improvement in SVT alignment. Also, analysis method was improved.
When applied to previous data of 23M $B\overline B$ events, the
result was $0.32\pm0.18$. But for the 2001 data of 9M, the result
was $0.83\pm 0.23$, which deviated by 1.8$\sigma$ from 1999-2000
result. On the other hand, Belle reported at LP01 the stunning
result of\cite{phi101}
\begin{equation}
 \sin2\phi_1 = 0.99 \pm 0.14 \pm 0.06. \ \ \ \ ({\rm Belle,\ 31.3M}\ B\overline B)
\label{eqn:phi101}
\end{equation}
Although at some variance, the combined result of 0.79 not only
established CPV in $B$ system beyond doubt. It is also quite
consistent with the 2004 HFAG average, showing the power of having
two B factories.
Both Belle and BaBar published details of their 2001 analysis,
while another round of analyses in 2002 gave
\begin{eqnarray}
 \sin2\beta &=& 0.741\pm0.067\pm0.034, \ \ \ \ ({\rm BaBar,\ 88M})
  \nonumber \\
 \sin2\phi_1 &=& 0.719\pm0.074\pm0.035, \ \ \ \ ({\rm Belle,\ 85M})
\label{eqn:phi102}
\end{eqnarray}
which are in rather good agreement. The case is fully settled.
We plot the measurements of
Eqs.~(\ref{eqn:phi1Osaka})--(\ref{eqn:phi102}) in Fig.~2.

\begin{figure}[t!]
\begin{center}
{\psfig{file=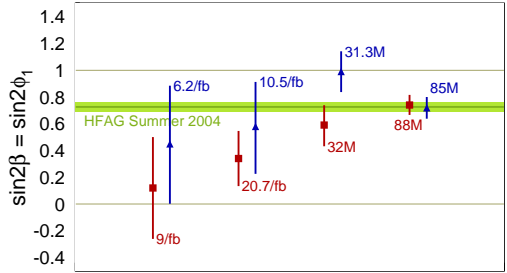,width=2.8in}} \vspace{-0.6cm} 
\end{center}
\caption{Measurements of $\sin2\phi_1/\sin2\beta$
         at the B factories. Square (triangle) is for BaBar (Belle).}
\end{figure}

Direct $CP$ asymmetry ${\cal A}_{J/\psi K_S}$ has also been
searched for and is found to be consistent with zero, confirming
the expectation in Eq.~(\ref{eqn:JpsiKs}). Polarization and triple
product correlations have also been studied in $B\to J/\psi K^*$
decays, where 3 partial waves are present. Although evidence is
found for final state interactions, no indication was found for
deviations from SM. The current average $\sin2\phi_1$ value of
Eq.~(\ref{eqn:phi104}) is in good agreement with a global
fit\cite{global} to $\epsilon_K$, $\Delta m_{B_d}$,
$|V_{ub}/V_{cb}|$ data and limit on $B_s$ mixing. Up to a four
fold ambiguity, the preferred value for $\phi_1$ in Fig.~1 is
$23.3^\circ\pm 1.6^\circ$.

The {\it raison d'\^etre} of the B factories was already completed
within two years of turning on, and $\sin2\phi_1$ is now a
precision measurement. No indication is found for deviation from
the KM picture of CPV.

\section{$B^0 \to \pi^+\pi^-$ and $\phi_2/\alpha$ Measurement}

In early times, it was thought that $\overline B^0\to \pi^+\pi^-$
decay would proceed by the $b\to u\bar ud$ tree diagram, hence the
decay amplitude ratio $\langle \pi^+\pi^-|S|\overline B^0\rangle /
\langle \pi^+\pi^-|S|B^0\rangle = V_{ub}/V_{ub}^* = e^{-2i\phi_3}$
(or $e^{-2i\gamma}$), so $\lambda_{\pi^+\pi^-} =
e^{-2i\phi_1}\,e^{-2i\phi_3} = e^{-2i(\pi-\phi_2)} =
e^{+2i\phi_2}$. One could thus measure $\sin2\phi_2$ via the
$\pi^+\pi^-$ mode. If $b\to u\bar ud$ tree dominance were true,
one would also expect ${\cal A}_{\pi^+\pi^-}$ to vanish.

This picture was already shattered by the CLEO
observation\cite{CLEOKpi} of $K^-\pi^+$ before $\pi^-\pi^+$. The
$b\to u\bar us$ tree process should be suppressed by
$|V_{us}/V_{ud}|^2\sim 1/20$ in rate w.r.t. the $b\to u\bar ud$
tree process. The CLEO observation clearly demonstrates that the
loop-induced ``penguin" $b\to s\bar qq$ process dominates
$\overline B^0\to K^-\pi^+$. With a $K^-\pi^+$ rate 4 times that
of $\pi^+\pi^-$ mode, we now expect the penguin amplitude to be of
order 30\% of the tree for $\overline B^0\to \pi^+\pi^-$ decay.
The penguin amplitude not only brings in weak phases, it could
introduce strong phases relative to the tree amplitude as well. It
is common practice to write\cite{GroRos}
\begin{equation}
 \lambda_{\pi^+\pi^-} = e^{+2i\phi_2} \,
                        \frac{T+P\,e^{+i\phi_3}e^{i\delta}}
                             {T+P\,e^{-i\phi_3}e^{i\delta}}
                      = \frac{e^{+i\phi_2}-\hat P\,e^{-i\phi_1}e^{i\delta}}
                             {e^{-i\phi_2}-\hat P\,e^{+i\phi_1}e^{i\delta}},
 \label{eqn:lpipi}
\end{equation}
where $T$ and $P$ are the magnitudes of the tree and penguin
amplitudes, and $\delta$ is their strong phase difference. One can
see that, if $\hat P\equiv P/T \rightarrow 0$, then
$\lambda_{\pi^+\pi^-} \rightarrow e^{+2i\phi_2}$, but for $P/T
\sim 0.3$, the extraction of $\phi_2$ becomes rather complicated.
The presence of $P$ and $\delta$ also bring in hadronic
uncertainties that are hard to deal with theoretically. As there
are more parameters than measurables, an isospin
analysis\cite{GroLon} involving $\pi^+\pi^0$ and $\pi^0\pi^0$
modes is necessary to fit for $P/T$ and $\delta$, in addition to
$\phi_2$ and $\phi_1$.

\subsection{${\cal S}_{\pi\pi}$, ${\cal A}_{\pi\pi}$ Measurements}

As if theoretical difficulties were not enough, Belle and BaBar
have not yet reached mutual agreement on their measurements of
${\cal S}_{\pi\pi}$ and ${\cal A}_{\pi\pi}$.

The first measurement was reported by BaBar in 2001 based on 33M
$B\overline B$ pairs,\cite{alpha01}
\begin{equation}
 {\cal S}_{\pi\pi} = +0.03^{+0.53}_{-0.56}\pm0.11,
  \ \ \
 {\cal C}_{\pi\pi} = -0.25^{+0.45}_{-0.47}\pm0.14, \ \ \ \ ({\rm BaBar,\ 33M})
\label{eqn:alpha01}
\end{equation}
which is a null measurement serving the purpose of establishing
technique. Then came the astonishing result\cite{phi202} from
Belle in 2002, based on $\sim 45$M $B\overline B$ pairs,
\begin{equation}
 {\cal S}_{\pi\pi} = -1.21^{+0.38+0.16}_{-0.27-0.13},
  \ \ \
 {\cal A}_{\pi\pi} = +0.94^{+0.25}_{-0.31}\pm0.09, \ \ \ \ ({\rm Belle,\ 45M})
\label{eqn:phi202}
\end{equation}
which is in strong contrast (note that ${\cal A}_f = -{\cal C}_f$;
see Eq.~(\ref{eqn:ACPt})) with Eq.~(\ref{eqn:alpha01}). Thus
commenced the controversy between BaBar and Belle on ${\cal
S}_{\pi\pi}$, ${\cal A}_{\pi\pi}$. We note that, although ${\cal
S}_{\pi\pi}^2 + {\cal A}_{\pi\pi}^2 \leq 1$ according to
Eq.~(\ref{eqn:SfAf}), because of dilution (mistag) factors in the
actual measurement of Eq.~(\ref{eqn:ACPt}), it is possible for
measured ${\cal S}_{\pi\pi}$, ${\cal A}_{\pi\pi}$ values to lie
outside the ``physical" region, especially if the CPV effect is
large.

The next round came summer 2002 from BaBar with more than twice
the data,\cite{alpha02}
\begin{equation}
 {\cal S}_{\pi\pi} = +0.02\pm0.34\pm0.05,
  \ \ \
 {\cal C}_{\pi\pi} = -0.30\pm0.25\pm0.04, \ \ ({\rm BaBar,\ 88M})
\label{eqn:alpha02}
\end{equation}
which confirmed their earlier result. This was followed by Belle
in early 2003,\cite{phi203}
\begin{equation}
 {\cal S}_{\pi\pi} = -1.23\pm0.41^{+0.08}_{-0.07},
  \ \ \
 {\cal A}_{\pi\pi} = +0.77\pm0.27\pm0.08, \ \ \ ({\rm Belle,\ 85M})
\label{eqn:phi203}
\end{equation}
which also confirmed their earlier result. The conflict continued.

At summer 2003 conferences, BaBar updated with the
result\cite{alpha03}
\begin{equation}
 {\cal S}_{\pi\pi} = -0.40\pm0.22\pm0.03,
  \ \ \
 {\cal C}_{\pi\pi} = -0.19\pm0.19\pm0.05, \ \ ({\rm BaBar,\ 122M})
\label{eqn:alpha03}
\end{equation}
which seems to move towards the Belle value. The reprocessing of
old 88M $B\overline B$ pair data gave
${\cal S}_{\pi\pi} = -0.252\pm0.27$,  ${\cal C}_{\pi\pi} =
-0.166\pm0.22$, while the newly added 34M data gave
${\cal S}_{\pi\pi} = -0.67\pm0.35$,  ${\cal C}_{\pi\pi} =
-0.33\pm0.34$. This result went unpublished, probably because
BaBar sought confirmation of the ``move" with more data.

In early 2004, Belle announced\cite{phi204} the observation of
large CPV in $B^0\to \pi^+\pi^-$,
\begin{equation}
 {\cal S}_{\pi\pi} = -1.00\pm0.21\pm0.07,
  \ \ \
 {\cal A}_{\pi\pi} = +0.58\pm0.15\pm0.07, \ \ ({\rm Belle,\ 152M})
\label{eqn:phi204}
\end{equation}
claiming a 5.2$\sigma$ effect w.r.t. ${\cal S}_{\pi\pi} = {\cal
A}_{\pi\pi} = 0$, and 3.2$\sigma$ evidence for DCPV (${\cal
A}_{\pi\pi} \neq 0$), regardless of ${\cal S}_{\pi\pi}$ value,
indicating the presence of strong phases. Note that the value of
${\cal S}_{\pi\pi}^2 + {\cal A}_{\pi\pi}^2$ now does touch the
``physical" boundary of 1.

BaBar updated with 227M $B\overline B$ pairs at summer 2004
conferences,\cite{alpha04}
\begin{equation}
 {\cal S}_{\pi\pi} = -0.30\pm0.17\pm0.03,
  \ \ \
 {\cal C}_{\pi\pi} = -0.09\pm0.15\pm0.04, \ \ ({\rm BaBar,\ 227M})
\label{eqn:alpha04}
\end{equation}
which, with almost doubling of summer 2003 data, moved slightly
back. Belle has just updated\cite{phi205} with full 2004 dataset
of 275M $B\overline B$ pairs, giving
\begin{equation}
 {\cal S}_{\pi\pi} = -0.67\pm0.16\pm0.06,
  \ \ \
 {\cal A}_{\pi\pi} = +0.56\pm0.12\pm0.06, \ \ ({\rm Belle,\ 275M})
\label{eqn:phi205}
\end{equation}
where ${\cal S}_{\pi\pi}$ shifted downwards by more than
1$\sigma$, and now the central value satisfies ${\cal
S}_{\pi\pi}^2 + {\cal A}_{\pi\pi}^2 \leq 1$. However, the conflict
between Belle and BaBar remains at $\sim 2.3\sigma$.
The results of Eqs. (\ref{eqn:alpha01})--(\ref{eqn:phi205}) are
plotted in Fig.~3.

\begin{figure}[t!]
\begin{center}
{\psfig{file=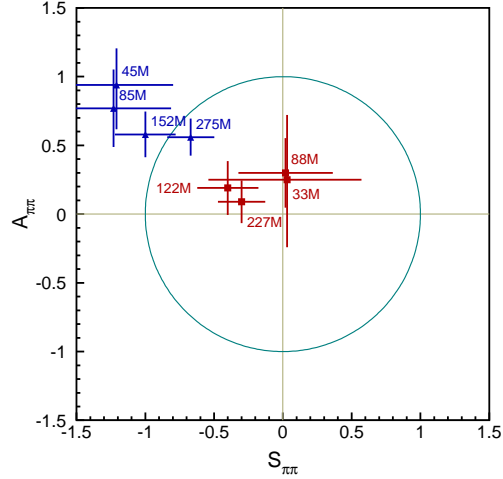,width=2.7in}} \vspace{-0.5cm} 
\end{center}
\caption{Measurements of ${\cal S}_{\pi\pi}$ and ${\cal
A}_{\pi\pi}\ (=-{\cal C}_{\pi\pi})$. Square (triangle) is for
BaBar (Belle).}
\end{figure}

It is not clear whether the deviation between Belle and BaBar on
$\pi^+\pi^-$ results is due to background, analysis method, or
statistical fluctuation.
The latest Belle analysis (275M) contains 2820 candidate events,
corresponding to $\sim 670$ $\pi^+\pi^-$ signal events, and 250
$K^{\pm}\pi^{\mp}$, 1900 continuum ($e^+e^-\to q\bar q$ where
$q=u$, $d$, $s$, $c$ quarks) background events in $M_{bc}$-$\Delta
E$ signal window. The latest analysis (227M) of BaBar, making a
multivariate, maximum likelihood and simultaneous fit for
$\pi^+\pi^-$ and $K^\pm\pi^\mp$ (and $K^+K^-$), is less
transparent: out of 68030 fitted events, $\sim 470\ \pi^+\pi^-$
events are extracted, together with $\sim 1600\ K^\pm\pi^\mp$
events. Compared to the $J/\psi K_S$ signal purity of over 97\%
and a larger effective rate, background is certainly much more
significant in the $\pi\pi$ analysis. From Fig.~3, however, it is
clear that there is some tendency of convergence between Belle and
BaBar as more data is added. But combining the results may not yet
be a good idea.

\subsection{$\pi^0\pi^0$, $\rho\pi/\rho\rho$ Modes and the $\phi_2/\alpha$ Program}

To extract $\phi_2/\alpha$ from $\pi\pi$ modes, an isospin
analysis\cite{GroLon} involving also the $\pi^+\pi^0$ and
$\pi^0\pi^0$ modes is needed. The $\pi^+\pi^0$ rate is relatively
well measured.\cite{HFAG} With 124M $B\overline B$ pairs, BaBar
found strong evidence\cite{pi0pi003} for the $\pi^0\pi^0$ mode
with rate at $(2.1 \pm 0.6 \pm 0.3) \times 10^{-6}$, roughly half
the $\pi^+\pi^-$ rate. This is much larger than factorization
expectations, but it is good for isospin analysis. Using 275M
$B\overline B$ pairs, Belle observed\cite{pi0pi0Belle} the
$\pi^0\pi^0$ mode at $(2.3^{+0.4+0.2}_{-0.5-0.3}) \times 10^{-6}$
with 5.8$\sigma$ significance. However, with 227M $B\overline B$
pairs, the BaBar number went down to\cite{pi0pi004}
$(1.17\pm0.32\pm0.10)\times 10^{-6}$ with 5.0$\sigma$
significance. A $2\sigma$ conflict exists, while the errors are
still too large for performing the $\phi_2$ program. And in any
case, there is an 8-fold ambiguity for $\phi_2$ determined this
way. The path to $\phi_2$ seems long and tortuous.

The $\rho\pi$ and $\rho\rho$ systems are the $VP$ and $VV$
counterparts of the $\pi\pi$ system, but clearly more complicated.
The $\rho\pi$ system cannot be a $CP$ eigenstate. In fact,
$\overline B^0 \to \rho^-\pi^+$ and $\rho^+\pi^-$ are both
possible. It has been suggested\cite{SnyQui} that, together with
$\overline B^0 \to \rho^0\pi^0$, a $t$-dependent Dalitz plot
analysis of $B^0/\overline B^0 \to \pi^+\pi^-\pi^0$ can in
principle determine $\alpha$ or $\phi_2$ without discrete
ambiguities, the latter resolved by the interference regions. Even
with the simplifying assumption of $(\rho\pi)^0$ dominance of the
Dalitz plot, this is a very difficult program. But BaBar has
pursued it, finding\cite{alpharhopi04} $\alpha =
(113^{+27}_{-17}\pm 6)^\circ$. We caution, however, that Belle and
BaBar do not yet agree on the strength of the $\rho^0\pi^0$ mode.
Given the disagreement in $\pi\pi$ results, we prefer to wait for
Belle to complete the Dalitz analysis. Belle has so
far\cite{phi2rhopi04} only analyzed by treating $\rho^\pm\pi^\mp$
as quasi-two-body, giving results agreeing with BaBar in general.
The combined\cite{HFAG} result of $A^{-+}_{\rho\pi} =
0.47^{+0.13}_{-0.14}$ gives over 4$\sigma$ evidence for DCPV in
the $\overline B^0 \to \rho^-\pi^+$ (but not in $\overline B^0 \to
\rho^+\pi^-$) mode,
%
which echoes the $\pi\pi$ mode. There seems to be strong phase
difference between tree and penguin amplitudes.

The $VV$ modes have 3 helicity amplitudes. BaBar has found 
the $B^0\to \rho^+\rho^-$ mode to be predominantly longitudinal,
hence is largely a $CP$ eigenstate. They also find, unlike the
$\pi\pi$ and $\rho\pi$ situation, a very small $\rho^0\rho^0$.
BaBar has therefore pursued the analysis vigorously. Their current
measurements of ${\cal S}_{\rho\rho}$ and ${\cal A}_{\rho\rho}$
are consistent with zero. Using isospin relations and their
results for $\rho^+\rho^0$ and $\rho^0\rho^0$ modes, BaBar
gives\cite{Roos} $\alpha = (96\pm10\pm4\pm11)^\circ$, where the
last error is due to penguin uncertainties. Belle, however, has
yet to give their results. Given that the $\rho\rho$ analysis is
more complicated than the $\pi\pi$ case, and BaBar and Belle are
in dispute on the latter, we feel it is premature to conclude.
We note, however, that $\alpha \sim 100^\circ$ is in good
agreement with the ``CKM fit"\cite{global} result not utilizing
CPV $B$ measurements.

\section{Penguin Dominant $b\to s$ Modes and New Physics}

As we have seen, penguin $b\to s\bar qq$ processes dominate over
the tree $b\to u\bar us$ process, enhancing e.g. the $\overline
B^0\to K^-\pi^+$ mode over the $\pi^-\pi^+$ mode by a factor of 4
in rate. The $b\to s$ penguins are induced by virtual loops
involving $u\bar u$, $c\bar c$ and $t\bar t$ quarks, which are
governed by the UT relation
\begin{equation}
 V_{us}V_{ub}^* + V_{cs}V_{cb}^* + V_{ts}V_{tb}^*  = 0.
\label{eqn:UTs}
\end{equation}
Unlike the UT relation of Eq.~(\ref{eqn:UT}) for $b\to d$
transitions, where all three terms are on equal footing, the first
term in Eq.~(\ref{eqn:UTs}), $|V_{us}V_{ub}^*| \cong
A\lambda^4\sqrt{\rho^2+\eta^2}$, is much smaller than the other
two terms $|V_{cs}V_{cb}^*| \cong |V_{ts}V_{tb}^*| \cong
A\lambda^2$. One has a rather collapsed UT compared to Fig.~1, and
$V_{cs}V_{cb}^*$, $V_{ts}V_{tb}^*$ are real to ${\cal
O}(\lambda^4)$. This implies that the decay amplitude is basically
real for penguin dominant modes. Thus, TCPV measurements in $b\to
s$ penguin dominant $CP$ eigenstates should give $\sin2\phi_1$,
just like the $b\to s(\bar cc)_{\rm charmonium}$ modes. This
constitutes a test of SM, and at the same time, any deviation
could indicate the presence of New Physics (NP). In the past
several years, the $B^0\to \phi K_S$ and $\eta^\prime K_S$ modes
have caused some sensation.

\subsection{Measurements in $B^0\to \phi K_S$ and $\eta^\prime K_S$}

The principles for TCPV study is the same as $J/\psi K_S$ and
$\pi^+\pi^-$ modes. The first measurement of a penguin dominant
$b\to s$ mode was performed by Belle for $B^0\to \eta^\prime K_S$
in 2002 with 45M $B\overline B$ pairs, giving\cite{etapKsBelle02}
\begin{equation}
 {\cal S}_{\eta^\prime K_S} = +0.28\pm0.55^{+0.07}_{-0.08},
  \ \ \
 {\cal A}_{\eta^\prime K_S} = +0.13\pm0.32^{+0.09}_{-0.06}, \ \ ({\rm Belle,\ 45M})
\label{eqn:etapKsBelle02}
\end{equation}
using $\eta^\prime\to \eta\pi^+\pi^-$ ($\eta\to \gamma\gamma$) and
$\rho^0\gamma$. This result is consistent with zero, but was soon
updated at ICHEP 2002, together with $\phi K_S$, to\cite{sqq02}
\begin{eqnarray}
 {\cal S}_{\eta^\prime K_S} = +0.71\pm0.37^{+0.05}_{-0.06},
  \ \;
 {\cal A}_{\eta^\prime K_S} &=& +0.26\pm0.22\pm0.03, \ \; ({\rm Belle,\ 85M})
\label{eqn:sqq02_etapKs} \\
 {\cal S}_{\phi K_S} = -0.73\pm0.64\pm0.22,
  \ \;
 {\cal A}_{\phi K_S} &=& -0.56\pm0.41\pm0.16, \ \; ({\rm Belle,\ 85M})
\label{eqn:sqq02_phiKs}
\end{eqnarray}
where ${\cal S}_{\eta^\prime K_S}$ became consistent with ${\cal
S}_{J/\psi K_S}$, but ${\cal S}_{\phi K_S}$ has the opposite sign!
This caused a sensation since BaBar also
reported\cite{phiKsBaBar02} a negative number,
\begin{equation}
 \ \ \ \ \ {\cal S}_{\phi K_S} = -0.19^{+0.52}_{-0.50}\pm0.09, 
 \hskip4.8cm ({\rm BaBar,\ 88M})
\label{eqn:phiKsBaBar02}
\end{equation}
which went unpublished. However, BaBar published the result for
$\eta^\prime K_S$,\cite{etapKsBaBar03}
\begin{equation}
\ \ {\cal S}_{\eta^\prime K_S} = +0.02\pm0.34\pm0.03,
  \
 {\cal C}_{\eta^\prime K_S} = +0.10\pm0.22\pm0.04, \ ({\rm BaBar,\; 89M})
\label{eqn:etapKsBaBar03}
\end{equation}
(again, ${\cal C} = - {\cal A}$ ) which seems more consistent with
Eq.~(\ref{eqn:etapKsBelle02}) than Eq.~(\ref{eqn:sqq02_etapKs}).

Summer 2003 was rather exciting. Belle updated with 152M,
giving\cite{sqq03}
\begin{eqnarray}
 {\cal S}_{\eta^\prime K_S} &=& +0.43\pm0.27\pm0.05,
  \
 {\cal A}_{\eta^\prime K_S} = -0.01\pm0.16\pm0.04, \; ({\rm Belle,\; 152M})
\label{eqn:sqq03_etapKs} \\
 {\cal S}_{\phi K_S} &=& -0.96\pm0.50^{+0.09}_{-0.11},
  \ \ \ \
 {\cal A}_{\phi K_S} = -0.15\pm0.29\pm0.07. \; ({\rm Belle,\; 152M})
\label{eqn:sqq03_phiKs}
\end{eqnarray}
While $\eta^\prime K_S$ mode moved down by 1$\sigma$, ${\cal
S}_{\phi K_S}$ became $\sim -1$ with 3.5$\sigma$ significance. But
the sign was no longer supported by BaBar, which reported with
114M,\cite{phiKsBaBar03}
\begin{equation}
 \ \ \ {\cal S}_{\phi K} = +0.47\pm0.34^{+0.08}_{-0.06},
  \ \;
 {\cal C}_{\phi K} = +0.01\pm0.33\pm0.10. \ \;
 ({\rm BaBar,\; 114M})
\label{eqn:phiKsBaBar03}
\end{equation}
Although the data increase is only 30\% or so, the move from
Eq.~(\ref{eqn:phiKsBaBar02}) is more than 1$\sigma$ because the
earlier dataset was reprocessed, and $\phi K_L$ events are also
incorporated. Once again one has disagreement between Belle and
BaBar, but the 2002 result and the 3.5$\sigma$ hint for potential
NP from Belle in 2003 stimulated many theory papers, mostly SUSY
models with down squark flavor violation, or $R$-parity violation.

The latest episode is no less dramatic. At ICHEP 2004, BaBar
gave\cite{etapKsBaBar04,phiKsBaBar04}
\begin{eqnarray}
 {\cal S}_{\eta^\prime K_S} &=& +0.30\pm0.14\pm0.02,
  \;
 {\cal C}_{\eta^\prime K_S} = -0.21\pm0.10\pm0.02, \, ({\rm BaBar,\, 232M})
\label{eqn:etapKsBaBar04} \\
 {\cal S}_{\phi K} &=& +0.50\pm0.25^{+0.07}_{-0.04},
  \ \ \ \ \
 {\cal C}_{\phi K} = \;\; 0.00\pm0.23\pm0.05, \ ({\rm BaBar,\; 227M})
\label{eqn:phiKsBaBar04}
\end{eqnarray}
while Belle gave\cite{sqq04}
\begin{eqnarray}
 {\cal S}_{\eta^\prime K_S} &=& +0.65\pm0.18\pm0.04,
  \;
 {\cal A}_{\eta^\prime K_S} = -0.19\pm0.11\pm0.05, \; ({\rm Belle,\; 275M})
\label{eqn:sqq04_etapKs} \\
 {\cal S}_{\phi K} &=& +0.06\pm0.33\pm0.09,
  \ \;
 {\cal A}_{\phi K} = +0.08\pm0.22\pm0.09. \; ({\rm Belle,\; 275M})
\label{eqn:sqq04_phiKs}
\end{eqnarray}
Except for ${\cal A}_{\phi K} \sim 0$, all three other
measurements are not in good agreement!

The Belle value for ${\cal S}_{\phi K}$ changed by 2.2$\sigma$,
shifting from $\sim -1$ in Eq.~(\ref{eqn:sqq03_phiKs}), to $\sim
0$ in Eq.~(\ref{eqn:sqq04_phiKs}). What happened was that the 123M
new data added in 2004 gave results with sign opposite to the
earlier 152M data. The new data was taken with the upgraded SVD2
silicon detector, which was installed in summer 2003. However, the
SVD2 resolution was studied with $B$ lifetime and mixing and is
well understood, while $\sin2\phi_1$ measured in $J/\psi K_{S/L}$
mode has good consistency between SVD2 and SVD1. Many other
systematics checks were also done. By Monte Carlo study of
pseudo-experiments, Belle concluded that there is 4.1\%
probability for the 2.2$\sigma$ shift.
Although the value is still 2$\sigma$ below $0.726$
(Eq.~(\ref{eqn:phi104})), given the large shift and the poor
agreement with the result from BaBar, which has been more stable
(though shifted 2002 $\to$ 2003 as commented earlier), one cannot
conclude whether there is signal for NP in $\phi K^0$ mode.

For the $\eta^\prime K_S$ mode, the indications from Belle and
BaBar are reversed compared to $\phi K_S$. The ${\cal
S}_{\eta^\prime K_S}$ value from BaBar is 3$\sigma$ below 0.726,
but the result from Belle is in good agreement with SM
expectations. Note also that, although the individual values for
${\cal C}_{\eta^\prime K_S}$ and ${\cal A}_{\eta^\prime K_S}$ are
not yet significant, they are again of opposite sign and are at
variance.
We conclude that one has to wait further to see whether there is
deviation from SM in TCPV in the $B^0\to \phi K^0$ and
$\eta^\prime K_S$ modes. The results for ${\cal S}_{\eta^\prime
K_S}$ and ${\cal S}_{\phi K_S}$ in
Eqs.~(\ref{eqn:etapKsBelle02})--(\ref{eqn:sqq04_phiKs}) are
plotted in Fig.~4.

\begin{figure}[t!]
\begin{center}
{\psfig{file=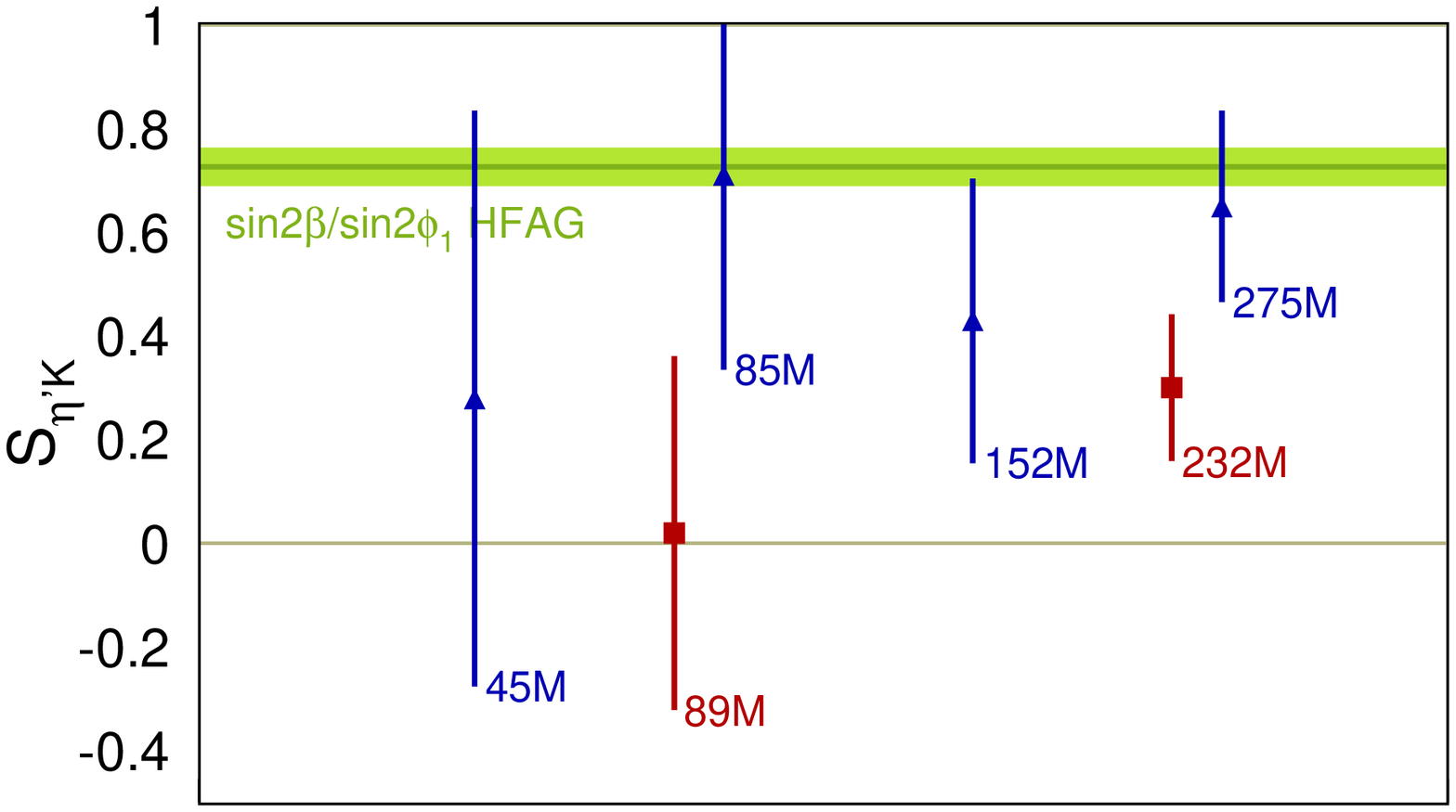,width=2.5in}\hspace{-0.3cm}
\psfig{file=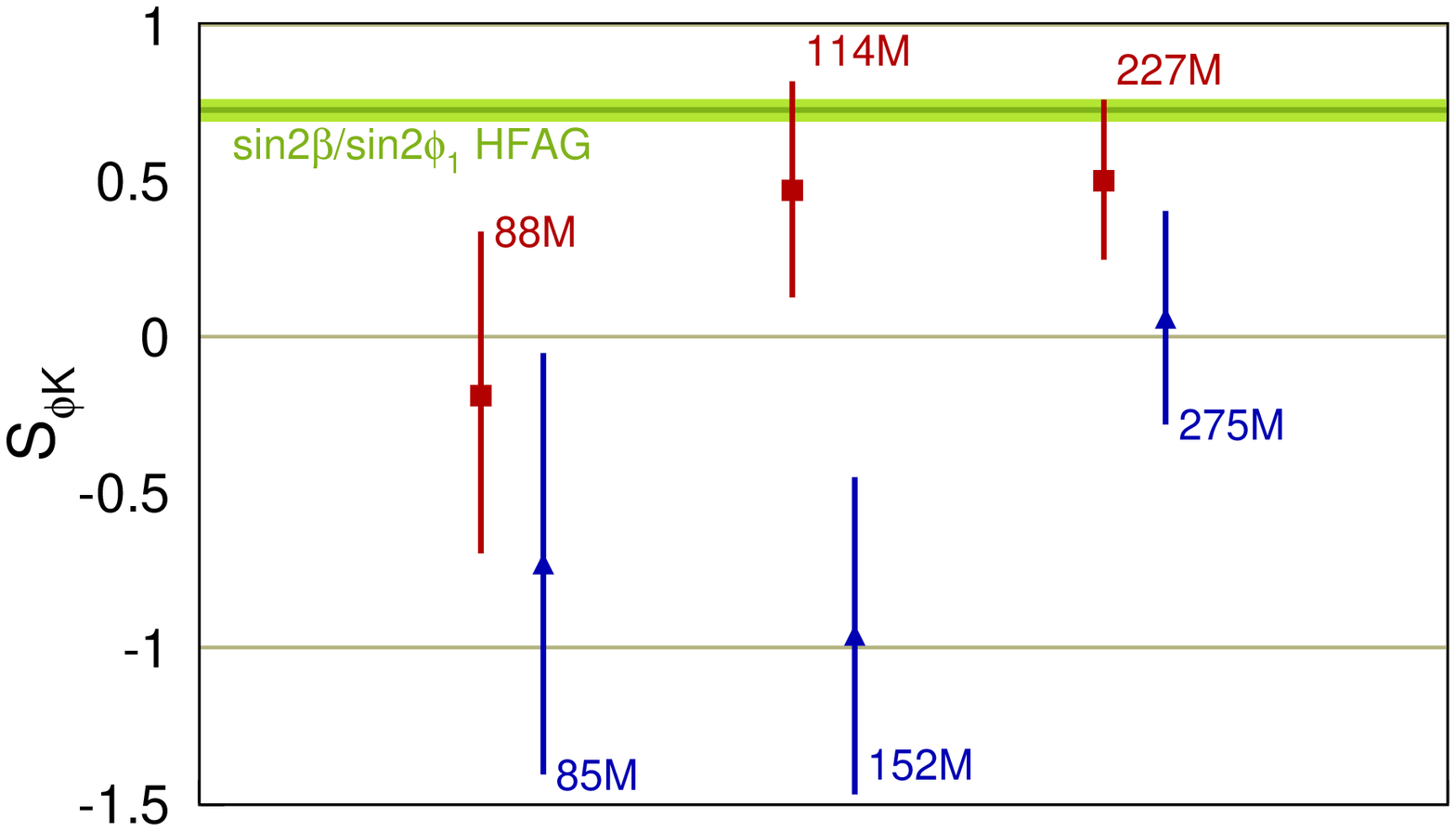,width=2.5in}}
 \vspace{-0.65cm} 
\end{center}
\caption{Measurements of ${\cal S}_{\eta^\prime K_S}$ and ${\cal
S}_{\phi K_S}$. Square (triangle) is for BaBar (Belle).}
\end{figure}

\subsection{Other Modes and Combined $b\to s$ Measurements}

A host of other penguin dominant $b\to s$ modes have also been
studied. The first such study\cite{sqq02} is $B^0\to K^+K^-K_S$
(excluding $\phi K_S$) by Belle, which has a large rate and was
found to be predominantly $CP$ even. By now one has
measurements\cite{HFAG} in $K^+K^-K_S$, $K_SK_SK_S$ ($CP$
eigenstate by angular momentum), $f_0(980)K_S$, $K_S\pi^0$, and
$\omega K_S$ modes. The latter is studied by Belle only,
disagreement exist in $K_SK_SK_S$ and $f_0(980)K_S$ modes, but the
effective $\sin2\phi_1/\sin2\beta$ value in $K^+K^-K_S$ and
$K_S\pi^0$ modes are 1$\sigma$ or more below 0.726. Although the
scatter among modes differ, the effective $\sin2\phi_1$ measured
in $b\to s$ penguins measured by Belle and BaBar are both
significantly below the charmonium result. The B factory average
is\cite{HFAG}
\begin{equation}
\sin2\phi_1(b\to s\bar qq) = 0.41\pm0.07,
\label{eqn:sin2phi1sqq}
\end{equation}
which is 3.8$\sigma$ below $\sin2\phi_1(b\to s\bar cc) =
0.726\pm0.037$. There seems to be some real effect in penguin
dominant $b\to s$ modes.

Besides the conflicts between Eqs.~(\ref{eqn:etapKsBaBar04}),
(\ref{eqn:phiKsBaBar04}) and Eqs.~(\ref{eqn:sqq04_etapKs}),
(\ref{eqn:sqq04_phiKs}), we remark that there are limitations for
what one can interpret from deviations in penguin dominant $b\to
s$ hadronic modes. While a large, definite effect in a single mode
such as $\phi K_S$ would clearly indicate NP, these modes suffer
from large hadronic uncertainties, such that the NP effect would
vary from mode to mode. So, whether $\phi K_S$ or $\eta^\prime
K_S$, or the combined effect in $b\to s\bar qq$, one does not gain
much more information by accumulating modes. It is difficult to
extract fundamental information of the underlying NP. What may be
more useful in the long run is $B^0\to K_S \pi^0\gamma$ decay.

The left-handedness of weak interactions imply $\overline B^0\to
K^*\gamma_L$, where $\gamma_L$ is a left-helicity photon; the
$\gamma_R$ component is suppressed by $m_s/m_b$ in amplitude. This
implies that, for $K^{*0}\to K_S\pi^0$, TCPV in $\overline B^0\to
K_S\pi^0\gamma$ $CP$ eigenstate is suppressed by $m_s/m_b$ hence
close to zero, because one needs the interference between
$\overline B^0$ and $B^0\to K_S\pi^0\gamma$ amplitudes with same
photon helicity, but one of which is always suppressed in SM. This
is therefore\cite{AGS} an excellent probe of NP that generates the
``wrong" helicity amplitude, whether the NP involves new CPV
phases or not.
%
This program seemed difficult because of poor vertex resolution
for the $K_S\pi^0\gamma$ final state. But in summer 2003, BaBar
showed\cite{Kspi0BaBar03} that TCPV could be measured in the
aforementioned $K_S\pi^0$ mode, utilizing a unique feature for the
B factories,
viz. that the $B$ direction is very close to the high energy beam
($e^-$ at present) direction.
This ``$K_S$ vertexing" technique can be applied to
$K_S\pi^0\gamma$, and has now been pursued by both BaBar and
Belle.
The beauty of studying TCPV in $B^0\to K^{*0}\,(K_S\pi^0)\,\gamma$
mode is that hadronic effects are largely in the $B\to K^*$ form
factor, which cancels in the TCPV measurement.
Thus, TCPV in $B^0\to K^{*0}\,(K_S\pi^0)\,\gamma$ mode provides a
fundamental measure of possible NP at a future high luminosity B
factory.

\section{The Quest for $\phi_3/\gamma$ in $B^\pm \to D^0 K^\pm$ Decay}

The measurement of $\phi_3/\gamma$ is known to be difficult. We
have now over $10^9$ $B$s, but we still have not extracted
$\phi_3$ convincingly with the so-called $DK$ method.

The idea of the $DK$ method is that $B^+\to DK^+$ decay can
proceed via two paths: $B^+\to \overline D^0K^+$ via $\bar b\to
\bar c\,u\bar s$, and $B^+\to D^0K^+$ via $\bar b\to \bar u\,c\bar
s$. While the former is Cabibbo suppressed and proportional to
$V_{cb}^*V_{us} \simeq A\lambda^3$, the latter is both doubly
Cabibbo suppressed and color suppressed, and proportional to
$V_{ub}^*V_{cs} \simeq A\lambda^3(\rho+i\eta)$. If $D^0$ and
$\overline D^0$ decay to a common final state, the two amplitudes
can interfere and probe the relative CPV phase $\phi_3$ between
$V_{ub}^*V_{cs}$ and $V_{cb}^*V_{us}$. The trouble is that one has
to measure very small branching ratios, controlled by the product
$r_B = |V_{ub}^*V_{cs}/V_{cb}^*V_{us}|\,F_{\rm cs}$, where $F_{\rm
cs}$ is the poorly known color suppression factor.

In the so-called GLW (or $D_{CP}K$) method,\cite{GLW} one studies
$\overline D^0/D^0$ decay to a $CP$ eigenstate, such as
$\pi^+\pi^-$. The interference effect is between a large and a
small amplitude, hence the effect is also small. In the so-called
ADS method,\cite{ADS} one studies $\overline D^0/D^0$ decay to a
flavor specific final state such as $K^-\pi^+$. In this way, one
brings down the $b\to c$ amplitude by selecting Cabibbo suppressed
$\overline D^0\to K^-\pi^+$ decay, thereby enhancing the
interference effect. Both methods have been studied by Belle and
BaBar, but so far they amount to limits on $r_B$ and are not yet
fruitful.

One interesting method was developed\cite{DalitzBelle0304,Giri}
recently involving three-body decays common to $\overline D^0$ and
$D^0$, such as $K_S\pi^+\pi^-$. Since this method uses information
from the Dalitz plot (including resonance phases), it is called
the $DK$ Dalitz plot analysis. Denote the $\overline D^0\to
K_S\pi^+\pi^-$ amplitude as $f(m_+^2,m_-^2)$, where $m_{\pm}^2 =
m_{K_S\pi^\pm}^2$ are the Dalitz variables. The corresponding
amplitude for $D^0\to K_S\pi^+\pi^-$ is therefore
$f(m_-^2,m_+^2)$. Thus, for $B^{\pm}\to (K_S\pi^+\pi^-)_D K^\pm$
decay, the amplitude is
\begin{equation}
f(m_{\pm}^2,m_{\mp}^2) + r_B
e^{i{\delta\pm\phi_3}}f(m_{\mp}^2,m_{\pm}^2),
\label{eqn:DKDalitz}
\end{equation}
where $\delta$ is the relative strong phase between the $b\to c$
and $b\to u$ amplitudes. Belle made the first study by modelling
$f(m_+^2,m_-^2)$ with known resonances, which was followed by
BaBar. The extracted results
are\cite{DalitzBelle0304,DalitzBaBar04}
\begin{eqnarray}
 \phi_3 = (77^{+17}_{-19}\pm13\pm11)^\circ, \ && \ ({\rm Belle,\ 152M})
\label{eqn:DKDalitzBelle04} \\
 \gamma = (70\pm26\pm10\pm10)^\circ, \ && \ ({\rm BaBar,\ 227M})
\label{eqn:DKDalitzBaBar04}
\end{eqnarray}
where the last error is from $f(m_+^2,m_-^2)$ modelling. Although
the $\phi_3/\gamma$ values are consistent, Belle finds a larger
$r_B$ than BaBar, which is partially reflected in the statistical
error. Belle has updated with 275M $B\overline B$ pairs,
finding\cite{DalitzBelle04}
\begin{eqnarray}
 \phi_3 = (68^{+14}_{-15}\pm13\pm11)^\circ. \ && \ ({\rm Belle,\ 275M})
\label{eqn:DKDalitzBelle05}
\end{eqnarray}
These results are consistent with CKM fit results.\cite{global}
Note that, with a much larger dataset, the model dependence can be
removed by using a binned fit\cite{Giri} over the Dalitz plot.

\section{Direct CP Violation in $B^0\to K^+\pi^-$}

Search for DCPV in $B$ system is important, since a variant of the
superweak model could be operative. It is remarkable that DCPV in
$B^0\to K^+\pi^-$ was observed already in 2004, just 3 years after
observation of mixing-dependent CPV.

Unlike mixing-dependent CPV where one needs decay time information
and tagging, the experimental study of DCPV is much simpler. They
are just counting experiments, and in the self-tagging modes such
as $K^\mp\pi^\pm$, one simply counts the difference between the
number of events in $K^-\pi^+$ vs. $K^+\pi^-$.

Indications for a negative DCPV in $B^0\to K^+\pi^-$ mode, defined
as
\begin{equation}
 {\cal A}_{K\pi} \equiv \frac
 {\Gamma(\overline B^0\to K^-\pi^+) - \Gamma(B^0\to K^+\pi^-)}
 {\Gamma(\overline B^0\to K^-\pi^+) + \Gamma(B^0\to K^+\pi^-)},
\label{eqn:AKpi}
\end{equation}
(basically the same definition as in Eq.~(\ref{eqn:SfAf})) have
been around for a couple of years. BaBar announced a
value\cite{AKpi1} with 4.2$\sigma$ significance just before ICHEP
2004, followed by Belle measurement\cite{AKpi2} with 3.9$\sigma$
significance. The results are,
\begin{eqnarray}
 {\cal A}_{K\pi} = -0.133\pm0.030\pm0.009, \
&& \ ({\rm BaBar,\ 227M})
\label{eqn:AKpiBaBar04} \\
 {\cal A}_{K\pi} = -0.101\pm0.025\pm0.005, \ && \ ({\rm Belle,\ 275M})
\label{eqn:AKpiBelle04}
\end{eqnarray}
combining to $-0.114\pm 0.020$ with 5.7$\sigma$ significance. This
establishes DCPV in $B$ system. The QCD factorization approach
predicted the opposite sign,\cite{QCDF} while the PQCD
factorization approach\cite{PQCDF} predicted the correct sign and
magnitude. Thus, the measurement has implications for the theory
of hadronic $B$ decays.

A tantalizing hint for new physics was uncovered by
Belle,\cite{AKpi2} and supported\cite{AKpi0BaBar04} by BaBar. By
isospin one expects ${\cal A}_{K\pi}$ in the $K^\mp\pi^\mp$ mode
and ${\cal A}_{K\pi^0}$ in the $K^\mp\pi^0$ mode to be very
similar. However, Belle and BaBar find
\begin{eqnarray}
 {\cal A}_{K\pi^0}
= +0.04\pm0.05\pm0.02, \ && \ ({\rm Belle,\ 275M})
\label{eqn:AKpi0Belle04} \\
 {\cal A}_{K\pi^0} = +0.06\pm0.06\pm0.01, \
&& \ ({\rm BaBar,\ 227M}) \label{eqn:AKpi0BaBar04}
\end{eqnarray}
combining to ${\cal A}_{K\pi^0} =0.049\pm0.040$, which deviates
from ${\cal A}_{K\pi} = -0.114\pm 0.020$ by 3.6$\sigma$. If this
result persists, it would imply NP in electroweak penguins
(mediated by $Z^0$ boson), which break isospin. However, the
previous BaBar measurement\cite{AKpi0BaBar03} with 88M $B\overline
B$ mesons gave ${\cal A}_{K\pi^0} = -0.09\pm0.09\pm0.01$. While
consistent with zero, the sign is opposite
Eq.~(\ref{eqn:AKpi0BaBar04}). Note further that, for $K^0\pi^\mp$
mode, we have\cite{AK0piBelle04,AK0piBaBar04}
\begin{eqnarray}
 {\cal A}_{K^0\pi}
= +0.05\pm0.05\pm0.01, \ && \ ({\rm Belle,\ 152M})
\label{eqn:AK0piBelle04} \\
 {\cal A}_{K^0\pi} = -0.087\pm0.046\pm0.010. \
&& \ ({\rm BaBar,\ 227M}) \label{eqn:AK0piBaBar04}
\end{eqnarray}
Though averaging\cite{HFAG} to $-0.02\pm0.04$, Belle and BaBar do
not agree in sign.
Thus, it is not yet clear whether ${\cal A}_{K\pi^0}$ and ${\cal
A}_{K^0\pi}$ have settled, although the ${\cal A}_{K\pi} - {\cal
A}_{K\pi^0}$ deviation should certainly be watched closely in the
near future.

\section{Discussion and Prospects}

We have left out many other highlights from the B factories, such
as the $B\to \phi K^*$ polarization puzzle, observation of $B\to
K^{(*)}\ell^+\ell^-$ and $X_s \ell^+\ell^-$, new hadron states,
etc. We chose to focus on significant CPV results from the B
factories.

It is clear that TCPV in $b\to s(\bar cc)_{\rm charmonium}$ modes
are now firmly established, with good agreement between Belle and
BaBar. What is surprising is that, while Belle and BaBar have each
made impressive TCPV measurements in $\pi^+\pi^-$, $\phi K_S$ and
$\eta^\prime K_S$ modes, agreement has not been reached in any of
these modes! Just compare Figs. 3 and 4 with Fig. 2. The
statistics may be still insufficient, and perhaps some algorithmic
improvements need to be made, since the charmless modes are not
background free.

We expect the $\pi^+\pi^-$ study to converge in a year or two, but
an isospin analysis may need a couple more years for $\pi^0\pi^0$
measurement to become more precise. Alternatively, if Belle
completes the $\rho\rho$ and/or $\rho\pi$ studies and concur with
the BaBar findings, then the B factories could claim the
measurement of $\alpha/\phi_2$ in a year or so. However, at this
point one cannot rule out further conflicts to develop.

The current 3.8$\sigma$ deviation between $\sin2\phi_1$ measured
from penguin $b\to s$ modes vs. $b\to s(\bar cc)_{\rm charmonium}$
is significant, and possibly hints at New Physics. But Belle and
BaBar disagree on the key $\phi K_S$ and $\eta^\prime K_S$ modes.
These two modes (as well as the higher statistics $K^+K^-K_S$
mode) may take several years to clear up as one needs a few times
more data. Other modes would have to wait even longer, and modes
like
$K_S\pi^0\gamma$ would probably have to await
the Super B factory with an order of magnitude or more increase in
luminosity.

The $\phi_3/\gamma$ measurement using $DK$ Dalitz analysis looks
promising.  In a few years it would become systematics limited,
and at the Super B factory one can use the model independent
binned fitting approach.

Direct CPV has been established in $B^0\to K^+\pi^-$ mode. We
expect a few more measurements to appear in next few years, such
as in $\pi^+\pi^-$, $\rho^\pm\pi^\mp$, and maybe $K^+\pi^0$,
$K^0\pi^+$ and $\eta K^+/\pi^+$. If the ${\cal A}_{K\pi} - {\cal
A}_{K\pi^0}$ difference persists, which may be known within a
year, then we may have New Physics in electroweak penguins.

We conclude that, before LHC starts to produce physics, we expect
$\alpha/\phi_2$ and $\phi_3/\gamma$ to be measured, and CKM
unitarity can be checked by direct measurement to some accuracy.
If New Physics effect is at the 20\% level or more for TCPV in
penguin $b\to s$ modes, it would be discovered. However, we may
know in a year or two whether we have New Physics in the
electroweak penguin.

\section*{Acknowledgments}

We thank K.F. Chen for help in graphics, and NSC for support.

\end{document}